\documentclass[twoside,twocolumn]{article}

\usepackage{blindtext} 

\usepackage[sc]{mathpazo} 
\usepackage[T1]{fontenc} 
\usepackage{microtype} 
\usepackage{color}

\usepackage[english]{babel} 

\usepackage[hmarginratio=1:1,top=10mm,right=10mm,columnsep=20pt]{geometry} 
\usepackage[hang, small,labelfont=bf,textfont=it,up]{caption} 
\usepackage{booktabs} 

\usepackage{lettrine} 

\usepackage{enumitem} 
\setlist[itemize]{noitemsep} 

\usepackage{abstract} 

\usepackage{titlesec} 
\renewcommand\thesection{\Roman{section}} 
\renewcommand\thesubsection{\roman{subsection}} 
\titleformat{\section}[block]{\large\scshape\centering}{\thesection.}{1em}{} 
\titleformat{\subsection}[block]{\large}{\thesubsection.}{1em}{} 

\usepackage{fancyhdr} 
\usepackage{titling} 

\usepackage{hyperref} 
\usepackage{subfig}
\usepackage{amsmath}
\usepackage{graphicx}

\pretitle{\begin{center}\Huge\bfseries} 
\posttitle{\end{center}} 
\title{Optimization of the post-crisis recovery plans in scale-free networks} 
\author{%
\textsc{Mohammad Bahrami$^1$}\\
\and 
\textsc{Narges Chinichian$^1$}\\
\and 
\textsc{Ali Hosseiny$^{1\*}$}\thanks{Corresponding author} \\[1ex] 
\and 
\textsc{Gholamreza Jafari$^{1,2}$}\\
\and 
\textsc{Marcel Ausloos$^{3,4}$}\\
\normalsize $^1$Department of Physics, Shahid Beheshti University, G.C. Evin, Tehran 19839, Iran \\ 
\normalsize$^2$Center for Network Science, Central European University, H-1051, Budapest, Hungary \\
\normalsize$^3$School of Business, College of Social Sciences, Arts, and Humanities, University of Leicester, Leicester, UK \\
\normalsize$^4$Department of Statistics and Econometrics, Bucharest University of Economic Studies, Bucharest, Romania \\
\normalsize $^*$ \href{mailto:al_hosseiny@sbu.ac.ir}{al\_hosseiny@sbu.ac.ir} 
}
\date{\today} 
\renewcommand{\maketitlehookd}{%
\begin{abstract}
\noindent 
General Motors or a local business, which one is it better to be stimulated in post-crisis recessions, when government stimulation is meant to overcome recessions? Due to the budget constraints, it is quite relevant to ask how government can increase the chance of economic recovery. One of the key elements to answer this question is to understand metastable features of crises in economic networks and their related hysteresis. The Ising model has been suggested for studying such features. In the homogenous networks, one needs at least a minimum budget, to force the network to switch its local equilibria, where such a minimum is independent of the network characteristics such as the average degree. In the scale free networks however, when the government aims to push the network to switch to another equilibrium, one may wonder which nodes are to be preferably stimulated  in order to minimize the cost. In this paper, it is shown that stimulation of high degree nodes costs less in general. It is also found that in scale free networks, the stimulation cost depends on the networks features such as its assortativity. Although we confine our study to the Ising model in order to tackle a problem in economics, our analysis shines lights on many other problems concerning stimulations of socio-economic systems where dynamical hysteresis appears.
\end{abstract}
} 

\begin{document}

\maketitle
\section{Introduction}
\lettrine[nindent=0em,lines=3]{I}n the aftermath of the 2007-2008 economic crisis, while the US government was going to stimulate the
 economy, some controversial issues had risen. 
For example, the debate about a recovery plan for the General Motors (GM) and Chrysler went up to the level 
of the US Senate. In that debate, some experts favored helping the big companies such as GM and Chrysler
 while   others favored small local businesses, see Stiglitz (2010)~\cite{stiglitz2010freefall} and
 references therein. 

To explain the problem more rigorously, we should recall that firms purchase products from their "neighbors" in 
the trade networks. Such a trade then results in positive correlations between activities of both  firms. Similar to the
ferromagnet systems, the positive correlation can result in a dynamical hysteresis  for the economic networks when in 
deep recessions one faces a global reduction in the activities of firms. In deep recession if the partners of a firm reduce their production and the firm in a different manner works with its maximum capacity, then there is a good chance that its products are not sold, but depreciate resulting in a loss. So, managers have no choice but keeping steps with their partners. Such a behavior can deepen economic harshness and may result in a long lasting depression. 

In Keynesian economics, governments are suggested to intervene in the market and purchase from the firms,
stimulating them to raise their activities in order to overcome recession. Due to the budget constraints,
it will be critical to find the best strategy for stimulation of a heterogeneous network, at least in a simple agent based model for firms.

The questions of \lq\lq{}whether one will better stimulate the recessed economy by helping the big
companies or the small ones\rq\rq{} is relevant due to the heterogeneous scale-free nature of 
economic networks. 
In other words, heterogeneity raises the question of finding the best \lq\lq{}strategy\rq\rq{} to help the system.
In a homogeneous network such as a regular lattice or a small-world Watts-Strogatz 
network~\cite{watts1998collective}, all nodes are connected to an almost similar number of neighbors, which have the same practical roles in the structure and have a similar priority for stimulation at crisis.

The strategy question opens a much more general problem of dealing with heterogeneous networks, 
where one cannot easily use a mean-field solution. In such cases depending on what is going to optimized, one needs to choose the best agents of the network to be stimulated. The low-degree nodes are easy to stimulate, while the high-degree nodes are more difficult but also more influential. So the answer to \lq\lq{} what is the best choice? \rq\rq{} will not in any way be trivial. To study this problem in a detailed way, we consider having several different scale-free networks with Ising spins on their nodes. 

The Ising model has been a proper choice to model a wide range of phenomena such as opinion 
dynamics~\cite{RevModPhys.81.591,opinion-dynamic,ausloos2008},
neural function simulations~\cite{hopfield1982neural}
and many other real-world systems, see for example~\cite{zhou, varela2015complex}. 

The Ising model can
 also be considered as a proper basis for our study,  - addressing the correlation of activities in
the economic network of firms~\cite{brock2001discrete,durlauf2010social,Hosseiny:2016}. 
Indeed, the Ising model has been suggested as a base to model the network of firms because the correlation between
the activities of firms can be encapsulated in the interaction between some spins. Since firms
trade with their neighbors in the network, when they increase/decrease productivity, they
automatically force their neighbors to increase/decrease productivity. In the simplest model, as a first approximation,  
one can imagine that managers  choose the firm level of activity from a binary choice being either the maximum or the minimum capacity of production. 

Since the trade is the way that firms interact with and influence  each other, in recessions, governments
can stimulate the economic networks via purchasing  products. The budget constraint however limits the 
governments choices;  therefore, which sectors or which classes of corporations are more appropriate choices for stimulation becomes a critical question.

If the activity of a network of firms is modeled by the Ising spins, then their response to the
external stimulation should be studied within the literature on the kinetic Ising model. The kinetic behavior of
the Ising model has been widely studied; for example, see~\cite{acharyya1995response,korniss2000dynamic,rikvold1994metastable,henkel2011non,
quantum-kinitic-ising}. Let it be recalled that if one imposes a magnetic field to an Ising system forcing it to switch its local equilibria, it takes some time for the
system to switch. Such a metastable behavior has been widely investigated for the regular networks.

The probability of switching between local equilibria as a function of the magnitude of the external field has been studied,  see for example~\cite{misra1997spin}. While the subject of such studies have been on regular networks, we are interested in the same type of study on scale free networks. 

 While regular networks   are homogeneous, scale free networks have an intrinsic heterogeneity leading to specifically interesting features \cite{scalefreeperc,scalefreenacher,scalefreehavlin2}. Moreover, this heterogeneity raises one question: should one discuss only the magnitude of the stimulating field? In fact or is it also relevant to ask: \lq\lq{} Where has the stimulus field to be imposed in the network? \rq\rq{}. The answer to these questions is the subject of this paper. Notice that although we have implemented the Ising model as a toy model of the network of firms, this study is interesting
also from a statistical physics point of view. It may as well shed light on a wide range of phenomena concerning metastabilities which occur in scale free networks.

In this work, we stimulate an Ising system to transfer it from one of its minima with all nodes first in a downward direction to the other minimum with almost all nodes in an upward direction. We will compare two general strategies: one starting our stimulation from the high
degree nodes (High-Degree-Stimulation strategy or HDS strategy) vs. another starting  from the low degree nodes (Low-Degree-Stimulation strategy or LDS strategy).
We investigate the amount of resources needed for each strategy.

We will observe that different strategies need different amount of hits, resources, or budget for success and will show that the gap between different strategies depends on some of the network characteristics. 

\section{Results}
In this Ising model of firms,  if firms work with maximum/minimum capacity, their state can be represented
by upward/downward spins. At the beginning of each session, managers of firms and corporations look at
their (collaborating) neighbors and subject to their level of activities decide to increase or decrease their
production level in the coming session, e.g., resulting in hiring or firing some employees.   

The chance for a manager's decision upon working with maximum/minimum capacity is  stochastic, following the probability rates   

\begin{eqnarray}
\begin{split}
P_{\uparrow}=\frac{\exp{-\frac{(N_\downarrow-N_\uparrow)J}{T}}}{\exp{-\frac{(N_\uparrow-N_\downarrow)J}{T}}+\exp{-\frac{(N_\downarrow-N_\downarrow)J}{T}}}
\\
\\
P_{\uparrow}=\frac{\exp{-\frac{(N_\uparrow-N_\downarrow)J}{T}}}{\exp{-\frac{(N_\uparrow-N_\downarrow)J}{T}}+\exp{-\frac{(N_\downarrow-N_\uparrow)J}{T}}},
\end{split}
\end{eqnarray}
where $P_{\uparrow/\downarrow}$ indicates the chance for working with maximum/minimum
capacity and $N_{\uparrow/\downarrow}$ indicates the number of neighbors which are working
with maximum/minimum capacity. The value of $J$ indicates the level of trade and
the value of $T$ weights how the manager is tied to her neighbors. If the value of $T$ is small
then the manager does not take risky actions in a sense that if the majority of their neighbors increase/decrease their
production level; then they also increase/decrease theirs with a high probability. 
\begin{figure}[!h]
\includegraphics[width=\columnwidth, height= 120mm]{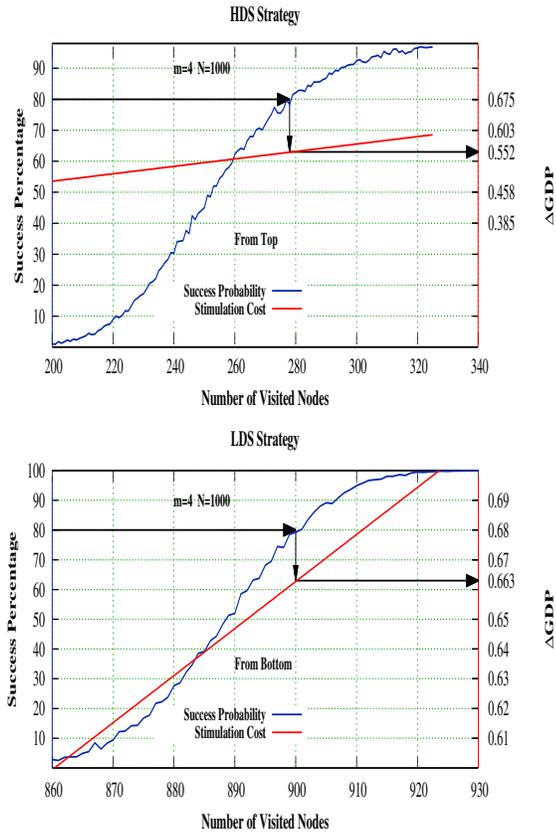}
\caption{ {\bf Different strategies.}\label{fig:Fig1}~(A) HDS strategy: In this simulation, all spins are set downward. 
We then start stimulation of high degree nodes. When $n$ highest degree nodes are stimulated with an 
upward magnetic field, then the system is relaxed to find the probability for a successful stimulation 
which changes the equilibrium of the network with the majority of nodes upward after relaxation. 
The horizontal axis shows the number of nodes quenched in each experiment and the blue curve is 
the success rate for such a quench state to switch to another global state. The red line shows the desired budget for 
such a quench, using an economic language. For a success rate of $80\%$ one needs a 
budget of $0.552\;\Delta GDP$ where $\Delta GDP$ is the gap for GDP between expansion and recession.
 \label{fig:Fig2}(B) LDS strategy: In this figure, the value of $n$ i on the horizontal axis indicates a quench state where $n$ lowest
  degree nodes are stimulated by a government.
   As it can be seen,  for having a success rate of $80\%$ one needs a budget equal to $0.663\Delta GDP$. Moreover, as it can be seen, for a stimulation with 80\% success chance, a stimulation
of low degree nodes costs about 20\% more than a stimulation of high degree nodes.}
\label{fig1}
\end{figure}
In the Ising model for reasonably small values of $T$ we observe symmetry breaking where if a big portion of firms decrease their production, then the system can stay there for a long time. 

In Keynesian economics government is suggested to compensate decline of neighbors to move economy from its metastable state. Thus, a government is suggested to start purchasing from the private parties, like  stimulating the system to move to its opposite local equilibrium. This action is similar to stimulation of   the spins with an external magnetic field. The government pays for the compensation caused by
the other firms' declining trades. As a result, the probabilities are modified as follows

\begin{eqnarray}\label{glabor}
\begin{split}
P_{\uparrow}=\frac{\exp{(-\frac{(N_\downarrow-N_\uparrow)J-GP}{T})}}
{\exp{(-\frac{(N_\uparrow-N_\downarrow)J-GP}{T})}
+\exp{(-\frac{(N_\downarrow-N_\uparrow)J+GP}{T})}}
\\
\\
P_{\downarrow}=\frac{\exp{(-\frac{(N_\uparrow-N_\downarrow)J+GP}{T})}}
{\exp{(-\frac{(N_\uparrow-N_\downarrow)J-GP}{T})}
+\exp{(-\frac{(N_\downarrow-N_\uparrow)J+GP}{T})}}
\end{split}
\end{eqnarray}
where $GP$ is a measure of the level of the purchase by government. 

As it is clear from the probabilities,  the government
purchase has some effect $GP$ is comparable with the trade between firms or strictly 
the value of $N_{\uparrow/\downarrow}J$.
Now, to address the response of the network to   stimulations, one should analyze the kinetic behavior
of the system under the change in size and strategy of stimulations.
To this aim, one first generates a preferential attachment scale-free  network (B-A Model~\cite{barabasi1999emergence}). Then, one sets all nodes to the downward direction indicating a situation where all firms work with minimum capacity. 

In the HDS strategy,  we start the  action by stimulating the high degree nodes. We consider a value for the number of nodes, $n$, and
impose a magnetic field on the $n$ high degree nodes updating them with the probabilities in Eq~\ref{glabor}. The stimulation imposed on each node is proportional to its degree

\begin{eqnarray}
GP_i=k_iJ.
\end{eqnarray}
After a number of nodes are stimulated, we let the system relax along sveral Monte Carlo steps in order to see if it changes its local equilibrium in such a way that the majority of firms starts working with their maximum capacity. 

\begin{eqnarray}
R_n = \sum_{i=1}^{n}GP_i.   
\end{eqnarray}
After a number of nodes are stimulated, we let the system relax in some Monte Carlo steps
to see if it changes its local equilibrium in the way that the majority of firms start working with their
maximum capacity. 

The resource needed for each strategy denoted by $R_n$ is the cumulative magnetic field imposed on
the nodes for the stimulation
\begin{eqnarray}
R_n = \sum_{i=1}^{n}GP_i.   
\end{eqnarray}
We repeat this for an ensemble of 1,000 experiments for all given values of $n$ and obtain
the success rate for such stimulation and its related resource. The result of the simulation is depicted in Fig~\ref{fig:Fig1}. In this figure, the blue curve shows the success rate for each value of $n$. 

For any given value of $n$, one  then measures the cumulative field, i.e. $R_n$, as the resource
needed for stimulation; in an economic language, it is the cost of the stimulation. 

\begin{figure}[!h]
\centering
\includegraphics[width=\columnwidth, height= 120mm]{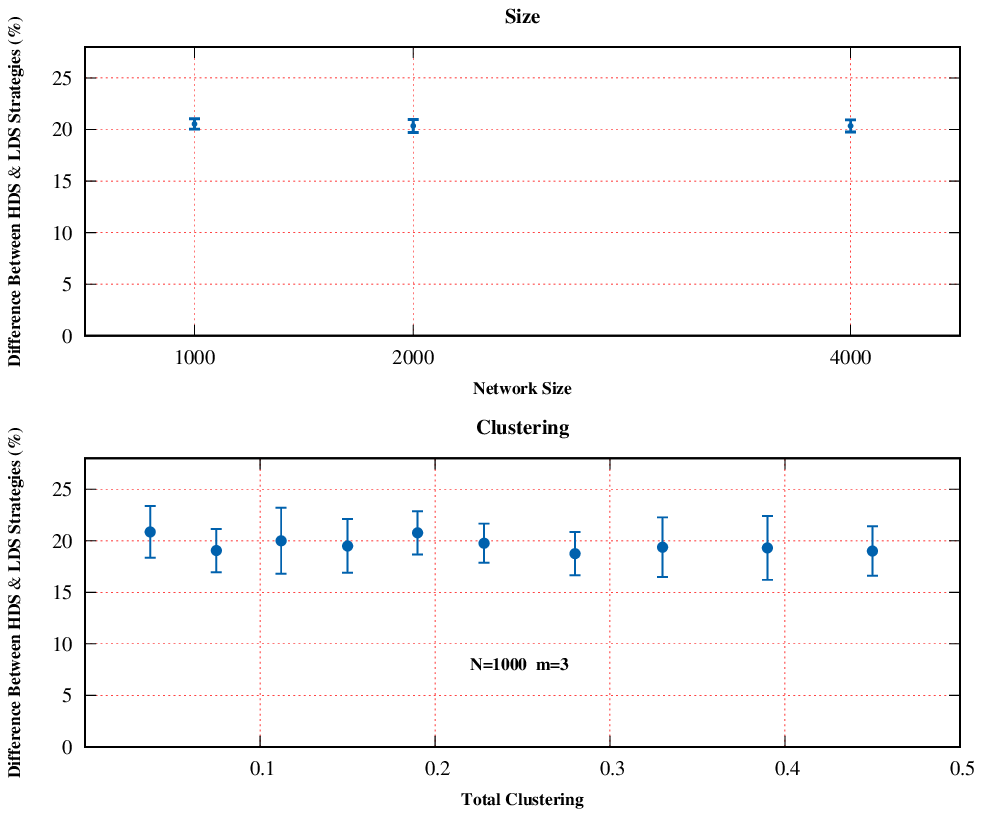}
\caption{ {\bf Evaluation of the impact of size and clustering on our results.}\\
\label{fig:size}(A) The difference for successful bills between HDS and LDS strategies for various sizes: As the size of the network grows from 1000 nodes to 4000 nodes, no significant difference is observed. In other words the gap between HDS and LDS strategies is independent of the size. \label{fig:clustering}(B) The impact of clustering: The difference between HDS and LDS stimulus bill is not influenced by clustering in the network.}
\label{fig:sizeandclus}
\end{figure}
 The red
curve shows the cost for each stimulation in $\Delta GDP$ units, where $\Delta GDP$ is the gap
for the gross domestic product $(GDP)$ between expansion and recession. This value is nothing else that
the total degree of the network or twice the number of links. The point is that in Eq~\ref{glabor} the parameters $N_{\uparrow\downarrow}J$ stand for the gap for the trade from each firm in expansion and recession. As a result, the gap for the GDP in expansion and recession periods, can be addressed by the total value of $N_iJ$, where $N_i$ is the degree of $i_{th}$. This is because the gap for GDP in expansion and recession can be reflected in the decline of trade between nodes as it has been encapsulated in $N_iJ$ for each node.
For more arguments concerning the relation between the role of GDP gap and  the average degree of nodes see~\cite{Hosseiny:2016}.

In our simulation we looked for stimulations which by a rate of $80\%$ success rate to change the global state of the spins. 
For this success rate in HDS strategy, we need to stimulate about 280 high degree nodes 
which is equal to spending  $0.552\;\Delta GDP$ in economic language or $R_n=1.104mNJ$ in the Ising language where $m$ is the average degree in the network. 

In LDS strategy everything is similar to HDS. We first set all spins downward and stimulate them.
 In this strategy however we stimulate low degree nodes. As shown in Fig~\ref{fig:Fig2}
to obtain a $80\%$ success rate, one needs a stimulation equal to $0.663\;\Delta GDP$.
This means that the cost of successful stimulation through the HDS strategy
is about $20\%$ less than the LDS strategy. 

The   difference in the outcomes of different strategies is significant. We however need to solidify the results. The first analysis to be done is the investigation
of the size dependency. We repeat  our simulations  for three different network sizes and
observed that our normalized results seem to be independent of the network size, see Fig~\ref{fig:size}.

Our first investigation appearing in Fig. \ref{fig:Fig1} is on a Barabasi-Albert (B-A) network.  A known feature of the B-A networks is that they have lower clustering coefficients than many real-world scale-free networks. To test the effect of
  the clustering coefficient, we change the clustering of our networks with the Holme-Kim method mentioned in the Methods section. It can be see that in Fig~\ref{fig:clustering}, changing the clustering coefficient has insignificant effect on the size of the gaps.

\begin{figure}[!h]
\centering
\includegraphics[width=\columnwidth]{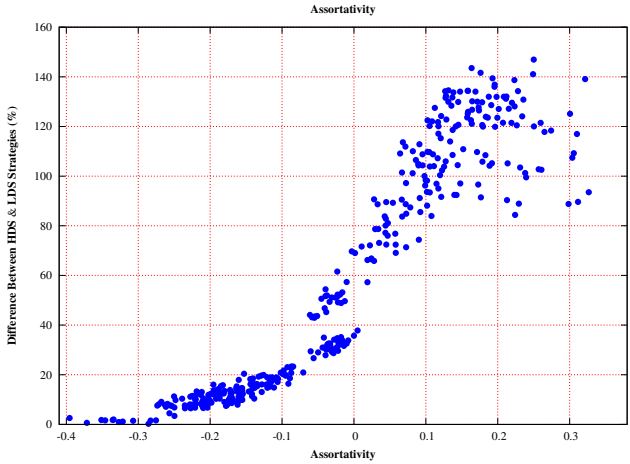}
\caption{ {\bf The impact of assortativity on the gap between different strategies.}~The gaps between the 
cost for HDS and LDS strategies grow as the assortativity of the networks grows.}
\label{fig:assort}
\end{figure}

Another feature of the real networks is their different assortativity structure. Assortativity is an important feature of the networks. It a measure for correlation of degrees. In other words it identifies if high degree nodes are preferably connected to high degree nodes, low degree nodes, or have no preferences. So, we perform another analysis
 to check the effect of assortativity on the stimulation strategies. Our simulation shows that, despite
  clustering, assortativity can significantly influence the gap between HDS and LDS strategies. When assortativity is increased in networks, the cost for LDS strategy remains unchanged. This is while the  HDS strategy becomes relatively cheaper and as a result, the gap between two strategies
     grows in size. For large values of the assortativity, the cost for LDS stimulation becomes more than
      twice the cost for HDS stimulation. Another observation is that the gap is saturated for large and
       small values of assortativity, see Fig~\ref{fig:assort}.

\section{Methods}
The Ising model is identified by the Hamiltonian
\begin{eqnarray}
H = -J \sum_{\langle i,j \rangle} s_i s_j -\sum_{j} h_j s_j,
\end{eqnarray}
where $s_i$ is the spin of the $i_{th}$ site, $J$ is the coupling constant, the notation $\langle i,j \rangle$ means that  the sum is over the nearest neighbor sites and $h_j$ indicates the stimulating external field  applied on the  $j_{th}$ node. 

To find the stimulation cost in Fig. \ref{fig:Fig2} all spins are set downward. Then in HDS strategy the high degree nodes are stimulated with a field equal to
\begin{eqnarray}
h=k_iJ.
\end{eqnarray}
After imposing a stimulation on a number of nodes, the system is relaxed to see if this stimulation is unsuccessful to push the network to switch to its other equilibrium where the majority of spins is upward. The temperature is set to $T=0.1T_c$. At equilibrium on this temperature $98\%$ of spins are downward. For the sake of simulation cost however in our analysis we set all spins downward which brings at least two percent systematic error. Such possible error however is small with respect to the cost gap  between HDS and LDS strategies.

For the study of the Ising dynamics on different networks, 
 the Glauber weight~\cite{glauber1963time} is used:
\begin{eqnarray}
W(s_i\to -s_i) = \frac{\exp^{-\beta\Delta E}}{1 + \exp^{-\beta\Delta E}},
\end{eqnarray}
where $\Delta E$ is the energy difference for the system 
changing the sign of $i_{th}$ unit and $\beta =\frac{1}{k_BT}$. 
The main targets of our study are scale-free networks (networks with power-law degree distribution). 

To have desired meta-stable states, the systems are considered below 
the critical temperature (in this case $T=0.1 T_c$). 
There are some analytical ways to find  
the critical temperature of Barabasi-Albert networks~\cite{bianconi2002mean,dorogovtsev2002ising,barabasi1999emergence}.

For the other networks introduced here, we used numerical methods and simulations to find
the critical temperature. \\
There are several known ways to produce scale-free networks~ 
\cite{barabasi1999emergence,caldarelli2002scale,scalefreehavlin,kumar2000stochastic,dangalchev2004generation}.

The methods used to produce or change them are listed below:  
\begin{itemize}
\item Barabasi-Albert Network was reconstructed using the algorithm mention 
in~\cite{barabasi1999emergence}; we generated the ensembles of B-A networks with total 
sizes of 1000, 2000 and 4000 nodes. The number of edges of new coming nodes ranged from 2 to 8 in different ensembles.
\item To change the "clustering" of our scale-free networks, we used the "triad formation" step by Holme
 and Kim~\cite{Holme:2002} 
keeping fixed the number of links, to be able to compare the results. 
\item  The assortativity~\cite{Newman:2011} of the scale-free networks is changed through the Brunet and
Sokolov reshuffling procedure~\cite{Xulvi:2004}.

\end{itemize}


\section{Discussion}

Huge studies have been devoted to the occurrence of crises and spread of shocks in economics. It has been shown for example that economies having more connections within the production networks may suffer intensive cascades of economic shocks \cite{battiston2012liaisons,contreras2014propagation}. 
In the homogeneous networks the central limit theorem 
rules out the chance for the systematic failure of the market due to the local fluctuations.
It has been however shown that unlike  the homogeneous networks, in the scale free networks, local fluctuations in the network may 
blow up and be the triggers for a crisis~\cite{acemoglu:2012}. This means that the structure of the economic networks can seriously influence their dynamic. 

While huge studies have been devoted to the occurrence and 
diffusion of the crisis~\cite{battiston2012liaisons,gatti2005new,shirazi2017non} in complexity economics ~\cite{arthur1999complexity,schweitzer2009economic,hosseiny2017geometrical,safdari2016picture,rotundo2010organization,rotundo2013complex,d2016complex,cerqueti2015review}, analyses of the responses of the networks to the recovery plans are lacking.

The Ising model as a base model to address dynamical properties of economic networks definitely simplifies the real world. While in this model heterogeneity is imposed only on the degree distribution, studies reveal that besides the degree distribution, size and influence of firms obey power law distributions \cite{gaffeo2003size,aoyama2010econophysics,rotundo2010organization,rotundo2009co}. Despite such simplifications, however, the model not only gives insights, but also leads to reasonable results.

In the deep recession of 2009, some experts including Nobel prize laureates, Paul Krugman and Joseph Stiglitz warned that only big stimulations could help  the economy moving toward a fast recovery \cite{krugman2012end,stiglitz2010freefall}. There was however  no  idea about how $big$ this stimulation should be.

In an analysis, through studying the hysteresis of the economic network along an Ising model, a threshold is suggested for the size of successful stimulations. This is shown that to overcome the recession the recovery bill should be bigger than such a threshold \cite{Hosseiny:2016}. For the recession of 2009 the model predicts the threshold for successful recovery plan for the case of the United State to be 650 billions of US $\$$. The recovery stimulus bill imposed by Obama administration was bigger than this threshold and successful.  Despite the United States, in the European Union stimulating bill was far below the threshold and failed to help a fast recovery, see \cite{Hosseiny:2016}.

In homogenous networks it is shown \cite{Hosseiny:2016} that the threshold for successful stimulation is universal and independent of the network properties. In this paper it was shown that not only response of the netwrok depends on its own structure, but also it depends on the strategies chosen by the government, i.e. the sectors where the stimulating bill is imposed on.

Our analysis shows that in general, it is more efficient to start stimulation from
high degree nodes. The resource gap between HDS and LDS strategies is independent of both the network size and its clustering coefficient. The gap between strategies however is influenced seriously by the assortativity value of the networks.
Networks with the highest assortativity show the largest gaps. Such results indicate that in order to obtaining some better estimate of  
the gap between HDS and LDS strategies, beside the degree distribution, we need to know other features of the studied network.

Back to the first question raised in the paper, our analysis suggests that the US government should have focused on big companies such as Chevrolet instead of small businesses. Due to the simplicity of the model, our findings might not be reliable for policy makers at this stage, nevertheles strongly suggesting further studies, investigations, and simulations based on real data. Dynamical hysteresis is not restricted to the Ising model. It exists in a wide range of systems where agents can influence each other. Actually positive correlations can lead to dynamical hysteresis. So, for more realistic models we expect dynamical hysteresis exists and only quantitative results are modified. An example is a model where firms can raise or decline production in a continouse level. Still for such model the dynamical hysteresis is observed and surprisingly the threshold for successful stimulation is close to the Ising model \cite{hosseiny2019hysteresis}.

One interesting discussion can occur if we consider the consequence of our hypothesis for the long run trends. In the Ising model there is a rich literature with respect to the response of the model to the external fields, see for example \cite{chakrabarti1999dynamic} and references therein. While the literature concerning homogenous networks is rich, such studies are lacking for heterogenous networks. Findings of the paper show that such studies will deepen our understanding of the metastable states in socio-economic systems.

\bibliographystyle{elsarticle-num}

\bibliography{main}

\end{document}